# Probing the evolution of electronic phase-coexistence in complex systems by terahertz radiation

G. L. Prajapati‡, Sarmistha Das‡, Rahul Dagar, V. Eswara Phanindra and D. S. Rana*

*Department of Physics, Indian Institute of Science Education and Research (IISER)*

*Bhopal, Madhya Pradesh-462066, India*

## Abstract

In complex oxides, the electrons under the influence of competing energetics are the cornerstone of coexistence (or phase-separation) of two or more electronic/magnetic phases in same structural configuration. Probing of growth and evolution of such phase-coexistence state is crucial to determine the correct mechanism of related phase-transition. Here, we demonstrate the combination of terahertz (THz) time-domain spectroscopy and DC transport as a novel strategy to probe the electronic phase-coexistence. This is demonstrated in disorder controlled phase-separated rare-earth nickelate thin films which exhibit metal-insulator transition in dc conductivity at around 180 K but lack this transition in terahertz (THz) dynamics conductivity down to low temperature. Such pronounced disparity exploits two extreme attributes: i) enormous sensitivity of THz radiation to a spatial range of its wavelength-compatible electronic inhomogeneities and ii) insensitivity to a range beyond the size of its wavelength. This feature is generic in nature (sans a photo-induced effect), depends solely on the size of insulating/metallic clusters and formulates a methodology with unique sensitivity to investigate electronic phase-coexistence and phase transition of any material system.

## 1. Introduction

Spatial coexistence of multiple phases in degenerate state is one of the basic features of *3d* transition metal oxides (TMOs). In such systems, the complex energetic interaction of lattice, charge, spin and orbital degrees of freedom hinder the onset of a sharp phase transition. Consequently, for a range of perturbation values around the transition point, the phase-coexistence of multiple phases occurs at nano to micrometer length scale. In this state, all the co-existing phases are fragile to an extent that a slight change in external conditions causes substantial change in their strength and spatial distributions. It is well established now that probing the growth and evolution of the phase-coexistence state upon varying the perturbation play a very crucial role in determining the correct mechanism of a phase transition. Some recent examples are as follows: (i) by imaging growth of metal-insulator coexistence in $VO_2$, Qazilbash *et al.*[1] revealed that the response of metallic puddles in phase-coexistence region is different from the pure metallic phase and the metal-insulator transition (MIT) in $VO_2$ is accompanied by mass divergence of charge carriers in the vicinity of transition temperature ($T_{MI}$); (ii) Post *et al.*[2] showed continuous domain wall formation (a second order transition) along with insulator cluster to metallic cluster transition (a first order transition) at the onset of MIT in rare earth nickelates ($RNiO_3$). This is a rare example of coexistence of first- and second-order phase transition in the entire class of TMOs; (iii) phase-coexistence in manganites suggest that the magnetoresistance behavior should be viewed as a percolation of metallic ferromagnetic domains into insulating phase[3]; and (iv) in high $T_C$ superconductors (HTSC), experiments revealed a partial gap in density of state exists for a range of temperatures above $T_C$ which provides microscopic basis for understanding of the fluctuating superconducting response above $T_C$ in hole-doped HTSC[4]. Such discoveries reported time to time continuously kept surprising as one goes deeper to understand the complexity in TMOs.

Visualization of the evolution of phase-coexistence and unveiling the underlying physics require development of novel experimental techniques and strategies. DC electrical transport, magnetometry, neutron diffraction and scattering, nuclear magnetic resonance, *etc.*, show substantial effect of evolution of phase-coexistence on the measured data which give an indirect clue of role of phase-coexistence in phase transitions. Direct spatial probes such as scanning tunneling microscope and spectroscopy and scanning near field infrared microscopy provide a real-time visualization of phase transitions and formation of phase-coexistence[1-4]. However, the probing of phase-coexistence state may be limited by sample preparation in desired conditions, need of sources which are very costly or need of specialized equipment and expertize. Thus, it is always in demand to develop experimental tools and methodology which are simple, cost effective, and widely available and can provide results in lesser time with high precision. Here, we present THz time-domain spectroscopy (THz-TDS) along with DC transport as a new methodology which can probe electronic phase-separation and its modulation with external perturbations.

In a DC transport measurement, the itinerant charge carriers move from one electrode to another due to applied voltage difference across the two electrodes. In this case, DC transport acts as a global probe since the entire volume of the sample participates in the electronic transport. In moving from one electrode to another electrode, the carriers suffer scattering and get localized into insulating clusters. Thus, in a given condition, the behavior of itinerant carriers in DC transport is the resultant of volume fractions of insulating/metallic clusters present in the system. On the contrary, THz-TDS is an optical, non-contact method to observe dynamic electronic transport in a sample. In this case, the THz pulse incidents only on a small area of the sample, thus, it works as a local probe. More importantly, within the probing area, the transmitted THz pulse will largely be affected by only those insulating/metallic clusters which are larger in size than the probing wavelength of THz pulse (200-1500 µm, in this study). The remaining smaller clusters will give an average effect. Thus, the response of the THz pulse not only depends on the volume fractions of insulating/metallic clusters but also depends on the number of those clusters which are of larger sizes than the probing wavelength of the THz pulse. We utilized these features to probe the electronic phase-coexistence in $RNiO_3$ system during the onset of MIT.

We fabricated a variety of nickelate thin films with varying cluster sizes in phase-coexistence region by incorporating different types of disorder into the films. The DC transport measurement continues to show the occurrence of temperature induce MIT in the films having different strength of disorder while in THz transport the transition weakens as the disorder in the film increases. Above a critical strength of disorder, the THz transport probes only metallic state of the film while the MIT is still observable in the DC transport. This feature in itself is surprising at first glance as one expects similar electronic transport at least qualitatively for a system when probed by two different techniques. We verified that this feature is not due to photo-induced effect and deviation in the two electronic transports occurs only in the films exhibiting electronic phase-coexistence below $T_{MI}$. Further, we established that this feature solely depends on the sizes of insulating/metallic clusters in phase-coexistence state, no matter how one achieves the phase-coexistence. Thus, this feature is generic in nature and true for electronic phase-coexistence in any material system. Our novel methodology introduced here, therefore, can serve as a pioneering tool to investigate electronic phase-coexistence present in any system and reveal correct mechanism of related phase transition.

## 2. Experimental Details:

In the present study, we prepared three sets of films: i) three films with varying strength of disorder exhibiting MIT of different strengths (set 1): $PrNiO_3$ (PNO) on $NdGaO_3$ (NGO) (100) [TF-1], $La_{0.5}Eu_{0.5}NiO_3$ (LENO) on NGO (100) [TF-2] and on $(LaAlO_3)_{0.3}(Sr_2TaAlO_6)_{0.7}$ (LSAT) (100) [TF-3] ii) two films without any MIT down to low temperature (set 2): LENO on $LaAlO_3$ (LAO) (100) [TF-4] and $LaNiO_3$ (LNO) on LAO (100) [TF-5] and iii) three films having moderate amount of cation disorder, one of them has mosaic disorder, all exhibiting MIT (set 3): $Pr_{0.5}Sm_{0.5}NiO_3$ (PSNO) on LAO (100) with thicknesses of 40 nm and 80 nm [TF-6 and TF-7, respectively] and PSNO on mosaic LAO (100) with thickness of 40 nm [TF-8] [see the Table 1]. All the films were grown using pulsed laser deposition (PLD) technique. Deposition parameters for different films can be found elsewhere.[5-8] DC resistivity of the films were measured using physical property measurement system (PPMS) in the range of 2-300 K. Measurements were performed in four-probe configuration to minimize the contact resistance. THz-TDS were carried out on these films in the temperature range of 5-300 K, using photoconductive antenna based spectrometer. THz data were collected by passing the THz waveform first through vacuum, then through the substrate and finally through the film deposited on very same substrate. To avoid any phase error, we processed the substrate in the same deposition condition prior to the measurements. After then, film was grown on the same substrate and the THz data were collected for the film.

## 3. Results and Discussion

### 3.1. Different electronic states below $T_{MI}$, probed by DC and THz transports:

A comparison of resistivities ($\rho$) of different films of set 1 [TF-1 to TF-3], measured in DC and THz (at 1 THz) transports, are shown in Figure 1 (a-c). THz dynamic resistivity was determined using the relation $1/\sigma_1(f) = \rho(f)$; where $\sigma_1(f)$ is the real part of complex THz optical conductivity ($\sigma^*(f)$) at frequency '$f$'. Clearly, for the film TF-1, both DC and THz resistivities exhibit MIT at ~ 130 K [Figure 1(a)]. Here, in the metallic region above $T_{MI}$, the DC and THz resistivities are nearly same, while below $T_{MI}$ at the onset of phase-coexistence state, the THz resistivity starts differing from the DC resistivity. The THz resistivity is about one order of magnitude lower than its DC counterpart at lowest measured temperature. This feature becomes more pronounced in the films TF-2 and TF-3 having larger disorder (will be discussed later). It is clearly seen that both of these films exhibit MIT in the DC transport while in THz transport, it gradually disappears [Figure 1(b-c)]. This is dominant to an extent that the THz resistivity in TF-3 exhibits only metallic behavior down to low temperatures. This is the first observation of a stark difference in electronic states probed by DC and THz transports. In general, one expects to observe similar electronic states in a given external condition, when a sample is probed by different techniques. THz-TDS being non-contact technique, may give slightly lower resistivity compare to DC resistivity of the film, the qualitative behavior of the film is still expected to be same in both the transport techniques. Hence, the probing of different electronic states below the $T_{MI}$ in DC and THz transport (insulating and metallic states, respectively) can raise several curiosities: what is the role of disorder in probing of different electronic states by the two different techniques? Is it related to disorder modulated phase-coexistence state or is photo-induced effect which melts the insulating state in THz probe giving rise to metallic state? We discuss such prospectives in detail as described below.

### 3.2. Role of Disorder

The thermal hysteresis of first order MIT is always susceptible to various types of disorder. In nickelates, such hysteresis as well as the electronic states above and below the MIT are very sensitive to the oxygen vacancies and the quenched disorder.[6, 9] In the films TF-1, TF-2 and TF-3, a combination of these two factors is responsible for inducing different levels of disorder. While TF-1 film is oxygen deficient, the disorder in TF-2 and TF-3 is induced by a combination of quenched disorder and oxygen deficiency. In TF-2 and TF-3, the quenched disorder manifests due to the large cation size mismatch of $La^{3+}$ (1.216 Å) and $Eu^{3+}$ (1.120 Å).[6] As larger tensile strain tends to develop larger oxygen vacancies in perovskite oxides, TF-3 [deposited on LSAT (100)] will be the more disordered film compare to TF-2 [deposited on NGO (001)].[10-12] Hence, expected disorder strength in ascending order is TF-1 to TF-3.

We further determined the strength of disorder quantitatively by analyzing complex THz conductivity spectra of the films. The complex THz conductivity ($\sigma^*$) vs THz frequency ($f$) spectra at different temperatures for TF-1, TF-2 and TF-3 are plotted in the Figure 1(d-f). It is clear from the spectra that the real component ($\sigma_1(f)$) of $\sigma^*$ for all the films deviate from Drude model as the zero-frequency peaks are shifted to the higher frequency region. Also, the imaginary component ($\sigma_2(f)$) is negative in some or whole frequency range at all temperatures. These features defy the conventional Drude theory of free charge carriers. Thus, the Drude-Smith (D-S) model was applied to explain the THz conductivity. As per D-S model, the $\sigma^*$ follows the relation: $\sigma^* = \frac{\epsilon_0 \omega_p^2 \tau}{1 - i\omega\tau}\left[1 + \frac{c}{1 - i\omega\tau}\right] - i\epsilon_0\omega(\epsilon_\infty - 1).$[13-15] Here, $\epsilon_0$ is permittivity of vacuum, $\epsilon_\infty$ is permittivity of the medium at higher frequency, $\omega_p$ is plasma frequency, $\tau$ is scattering time of the charge carriers and $c$ is persistence of velocity parameter which accounts the strength of disorder present in the system. Larger is the (negative) value of '$c$', stronger is the disorder. As shown in Figure 1(d-f), both $\sigma_1$ and $\sigma_2$ were simultaneously well-fitted to the D-S model. At 5 K, the disorder parameter '$c$' was determined as -0.47, -0.69 and -0.71 for TF-1, TF-2 and TF-3 films, respectively. Hence, order of the strength of disorder is same as we inferred qualitatively in previous section. Increasing disorder naturally increases electronic inhomogeneity in the system, thus strengthens the phase-coexistence state. This suggests that weakening of MIT in THz transport increases upon increasing the strength of phase-coexistence state which finally results in completely metallic state for the TF-3 film.

### 3.3 Phase-coexistence as the origin of probing different electronic states:

To confirm the phase-coexistence as the origin of the occurrence of different electronic states below $T_{MI}$ when probed by the DC and THz transports, it requires to compare DC and THz resistivities of nickelate films sans phase-coexistence. Therefore, two compressive films LaNiO$_3$ and La$_{0.5}$Eu$_{0.5}$NiO$_3$ on LaAlO$_3$ (100) substrate (*i.e.*, TF-4 and TF-5 films, respectively [set 2]), both are metallic down to 5 K, were investigated. Temperature dependences of DC and THz resistivities of these films are plotted in Figure 2(a). It is clearly seen that both of the films not only exhibit metallic behavior down to low temperature but also the magnitude of DC and THz resistivities are nearly the same. Occurrence of same electronic state in both the probe supports the idea that the observed disparity between the electronic states below the $T_{MI}$, probed by DC and THz transport in the TF-1, 2 and 3 films, is related to disorder modulated phase-coexistence state.

To further confirm the conclusion, it is imperative to consider the effect of photo-induced excitations in the films. This is because the possibility of charge-density-wave (CDW) formation in the insulating state of nickelates.[16-17] In general, the CDW energy gap lies in the range of 1-5 meV. Hence, the THz photons are capable to close this gap by melting the charge ordered insulating state and transition it into a metallic state with a vivid resonance behavior in the $\sigma_1(f)$ spectra.[16-17] However, no CDW like resonance absorption peak was observed in THz conductivity-frequency spectra [Figure 1(d-f)]. Other possibility is closing of charge transfer gap between O-$2p$ and Ni-$3d$ upper Hubbard bands due to photo-excitation. The charge transfer gap in nickelates lies in the energy range of ~1-2 eV while energy range of used THz band is ~0.8-6 meV, nearly three order of magnitude lower than the charge transfer gap. Still to discard this possibility, we performed THz measurements with three different power strengths of THz radiation beam on the TF-3 film as shown in Figure 2(b). We see that both $\sigma_1$ and $\sigma_2$ nearly overlap for all of the three cases. These results completely rule out any assumed photo-induced effect for the occurrence of metallic state in THz transport.[18]

### 3.4. Interplay between phase-coexistence and probing wavelengths of THz band:

Now, we attempt to draw a clear picture of the role of phase-coexistence and inherent attributes of the optical THz-TDS technique in probing metallic state below $T_{MI}$, in THz transport. To explain this feature, we invoke the attribute of the distribution of wavelengths in THz optical spectra used in the measurements. A schematic to visualize this effect is presented in the Figure 2(c). It depicts that, below the $T_{MI}$, the phase-coexistence state responds distinctly to DC and THz spectroscopic probing techniques. As shown in Figure 2(c), the carriers accelerated by DC field travel between the electrodes in four-probe geometry. These carriers will suffer scattering and get localized in insulating clusters with decreasing temperature, thus, resulting MIT in DC transport. In this picture, both the size and continuity of the insulating/metallic clusters are important aspects for the manifestation of MIT.[19] The THz conductivity, in contrast, involves non-contact mode measurement and is obtained by the optical attributes such as wavelength of the probing THz radiation. In present study, we extracted THz dynamic conductivity in the wavelength range of ~1500 – 200 μm (~0.2-1.4 THz). This implies that if the size of either of the insulating or metallic clusters is less than the wavelength of the THz spectrum then this technique will be insensitive to the existence of such domains/clusters and an average effect will be reflected in the measured outcome. This may be schematically visualized in Figure 2(c). Thus, the THz radiation with different wavelength/frequency components while transmitted through thin films (TF-1 – TF-3), will probe only those electronic clusters which are either compatible to or larger than the size of its wavelength. It is reiterated that TF-1 is the least disordered and TF-3 is the highest disordered. This implies that in TF-1, the insulating clusters below MIT is strong, long-ranged, and large-sized in comparison to other films. In this film, probing THz radiation will encounter insulating phases/clusters with sizes larger than its all wavelength components. Hence, the THz resistivity shall bear the signature of insulating phase and exhibit the MIT similar to that of the DC resistivity. In the case of the most disordered TF-3 film, the size of metallic clusters outgrows the insulating clusters below the MIT. Consequently, the THz wavelength becomes more comparable and sensitive to the large-sized metallic clusters and insensitive to the presence of the smaller sized insulating fractions. Thus, the THz conductivity remains metallic sans any MIT down to low temperatures. The TF-2 film with moderate disorder clearly shows the THz conductivity seeming to manifest a behavior which is intermediate of least disordered TF-1 and largest disordered TF-3. Lai *et al.*, in a similar context, suggested the superiority of AC current imaging in local limit with respect to the DC measurement for analyzing the local resistivity of the materials having phase-coexistence state.[20] In this framework, the THz spectroscopy proves to be even advanced non-contact tool to probe ac conductivity in local limit.

We further note that the THz and DC resistivities exactly match with each other for TF-4 and TF-5 films which lack phase-coexistence state and exhibit metallic state down to low temperatures. Thus, this study reveals the extraordinary sensitivity of the THz technique to the phase-coexistence; making it an indispensable technique for multi-faceted studies of complex systems in which various competing interactions define the evolution of exotic properties. From the data so far presented here, we conclude that controlling the electronic phase-coexistence via introduction of controlled amount of disorder leads to probing of completely metallic state below $T_{MI}$ in the THz transport, owing to the compatibility of the THz wavelength with spatial dimensions of insulating/metallic clusters. To check the generic nature of this feature whether it can be realized in other nickelate systems having other kinds of disorder, we investigated another set of nickelate films (set 3), one of them having mosaic disorder as described in next sections.

### 3.5. Generic nature of the response of phase-coexistence to THz-TDS:

We investigated transport properties of TF-6 and TF-7 thin film samples having a smaller cation mismatch ($Pr_{0.5}Sm_{0.5}NiO_3$: $r_{Pr^{3+}} = 1.179$ Å and $r_{Sm^{3+}} = 1.132$ Å) compare to $La_{0.5}Eu_{0.5}NiO_3$ films which allows a better control of quenched disorder. The THz resistivity at 1 THz, both in cooling and heating protocols, for TF-6 and TF-7 films are plotted along with DC resistivity in Figure 3 (a-b). In heating protocol, the THz resistivity exhibits a subtle MIT similar to DC resistivity; however, the THz resistivity is less than the DC resistivity at all the temperatures. In contrast, the THz resistivity in cooling protocol first exhibits MIT on cooling down to 120 K below which the metallic phase reappears upon further cooling (shaded region in Figure 3). This behavior is consistent at all the THz frequencies as shown in Figure 3 (c-d). The sudden fall of the THz resistivity curves in cooling protocol at around 100 K suggests the re-ordering/re-crystallization of the charge-ordered insulating state and occurrence of kinetically arrested phase transition.[21-22] This implies spatial re-distribution of insulating and metallic clusters as well as increase in the size of metallic clusters in phase-coexistence state below 100 K. One should note that this feature is not detected by the DC transport and the reason again lies in the compatibility of the THz wavelength with the size of insulating/metallic clusters. At the onset of MIT in these films, the insulating phase grows and then weakens at the cost of the growth of metallic phase; thus suggesting the re-distribution in the size of insulating clusters having either same or smaller size relative to the wavelength distribution of THz wave.

We increased further the strength of disorder in $Pr_{0.5}Sm_{0.5}NiO_3$ film by growing it on a mosaic substrate [TF-8 film]. We call it mosaic disorder, a disorder which is translated into the film through the mosaic structure of the substrate.[8] Figure 4 shows the comparison of DC and THz resistivities of TF-8 film. Here too, it is evident that the DC transport shows metal-insulator transition while the THz transport shows only metallic phase over the whole temperature range. This confirms the generic nature of the response of the electronic phase-coexistence state to THz-TDS. All types of established method to induce disorder in thin film motif yield the similar result; *i.e.*, strengthening of the metallic phase in the THz transport. We expect that this feature should be true for any system exhibiting electronic phase-coexistence, on modulating the insulating/metallic cluster sizes, for example in $VO_2$.

### 4. Conclusions

To conclude, we demonstrated a novel methodology *i.e.*, THz-TDS along with DC transport to probe electronic phase-coexistence. Upon modulating the strength of and insulating/metallic cluster sizes in phase-coexistence state via introducing controlled amount of disorder, DC transport still probes insulating state while THz-TDS probes metallic state below the $T_{MI}$. This study brings out two unique capabilities and applications comprising implementation of THz spectroscopy in complex materials prone to phase-coexistence, summed up as follows; i) this technique shows extraordinary sensitivity in unambiguous characterization of the spatial dimensions of electronic phase-coexistence state when the probing wavelength of THz radiation is comparable to the size of insulating/metallic clusters, thus, bringing a large variety of materials under the characterization ambit of this technique, and ii) insensitivity to such clusters of dimensions much smaller than the THz wavelength, thus, implying the capability of this technique to overlook a commonly prevalent clusters of smaller dimensions (in range of nm to few µm). While the former attribute is a novel characterization feature, the latter offers a new platform of device fabrication in the field of spintronics, involving conductivity modulation. We hope our novel methodology will be fruitful in future in characterizing electronic phase-coexistence in a variety of systems and determining mechanisms of related phase-transitions.

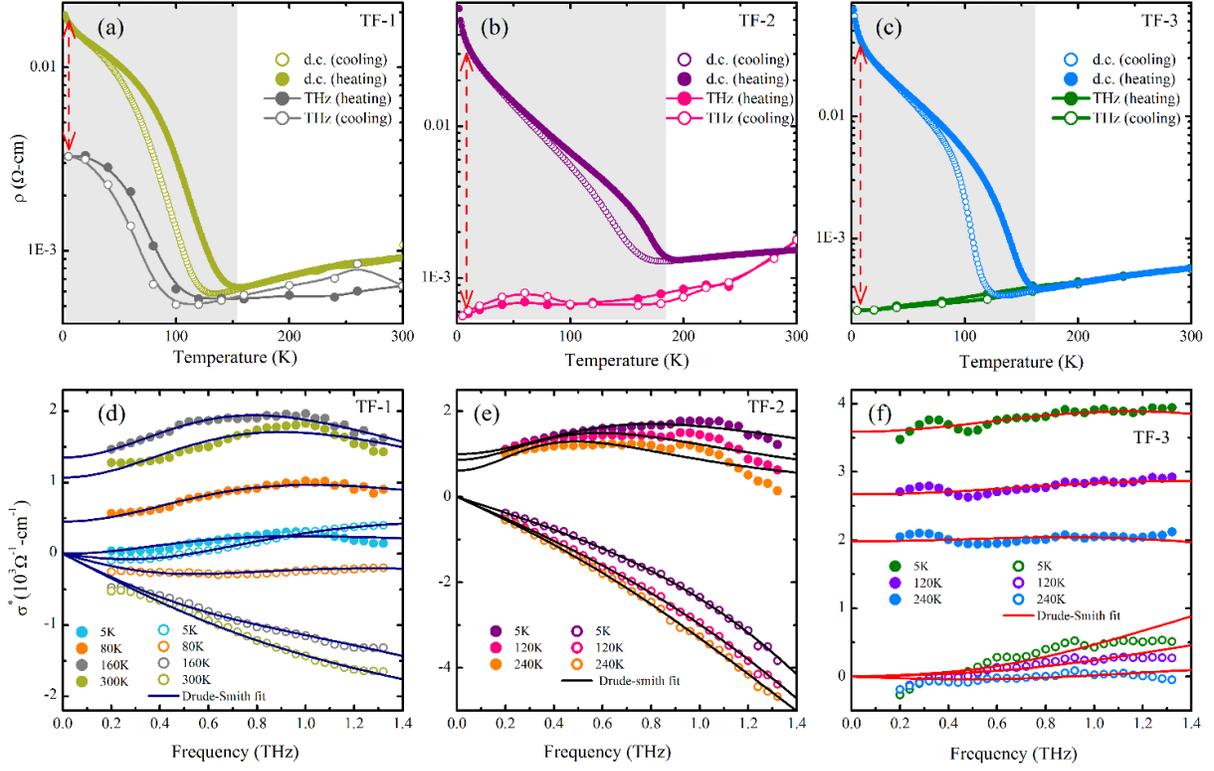

**Figure 1.** The comparative representation of DC and THz resistivities as a function of temperature for (a) TF-1, (b) TF-2 and (c) TF-3. The shaded region indicates the prominent difference between DC and THz resistivities below the $T_{MI}$. The $\sigma^*$ vs f spectra for (d) TF-1, (e) TF-2 and (f) TF-3. The filled and void symbols represent the real ($\sigma_1$) and imaginary ($\sigma_2$) component of $\sigma^*$, respectively. The solid lines are the fitted curves for Drude-Smith model.

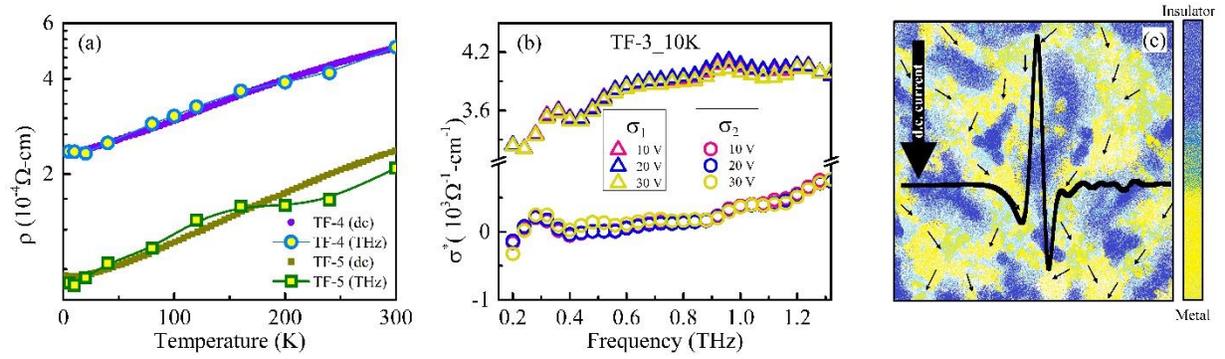

**Figure 2.** (a) The comparative representation of DC and THz resistivities as a function of temperature for TF-4 and TF-5. (b) $\sigma^*$ vs f spectra with THz emitter bias voltage variation for TF-3 at 10 K. (c) The schematic representation of phase-separated state consisting of insulating and metallic clusters. The small black arrows depict how the charge carriers get scattered in insulating clusters or trapped in defects in DC transport, whereas the THz waveform can able to give nearly similar average transmission in local limit.

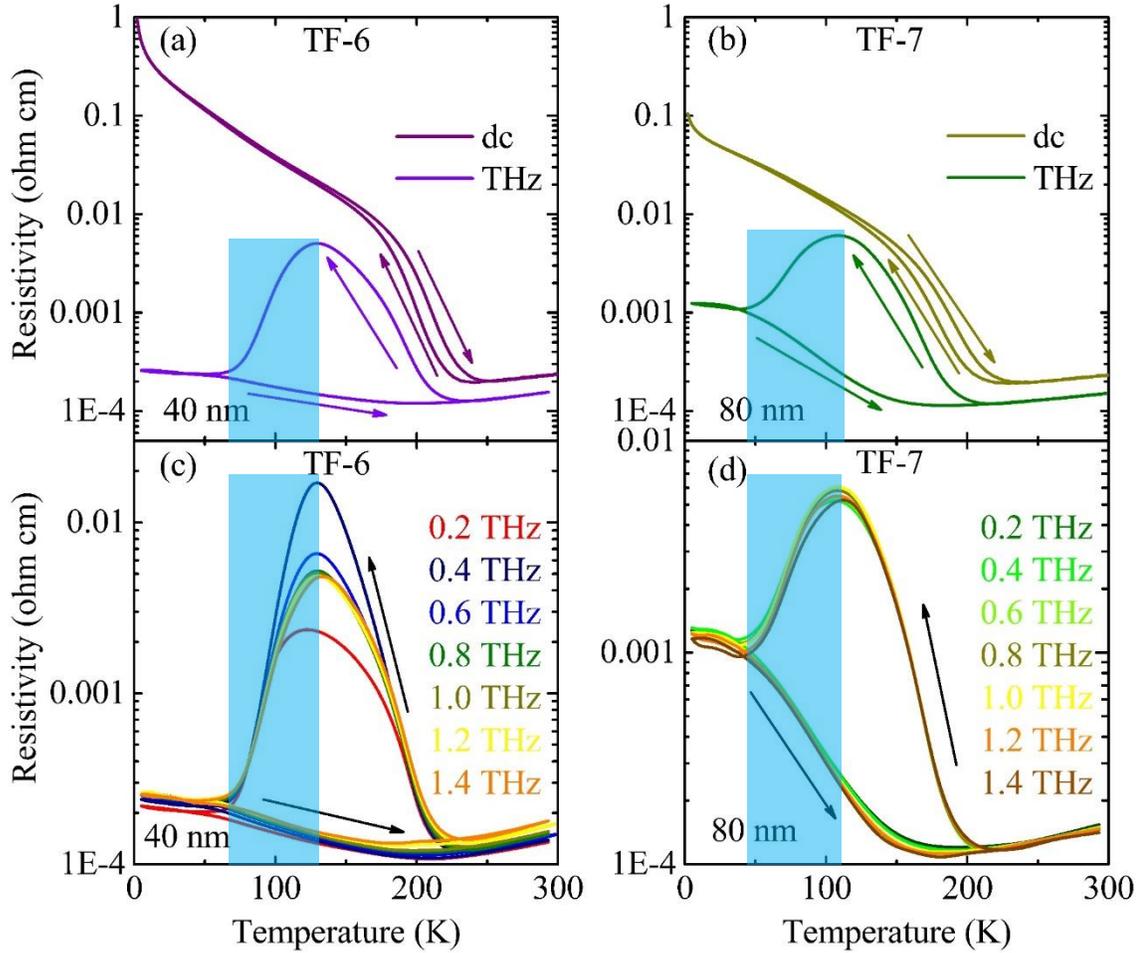

**Figure 3.** (a) and (b) Comparison of temperature-dependent resistivities obtained through DC transport and THz-TDS at 1 THz. Arrows show heating and cooling sequences. The shaded region in THz conductivity shows kinetically arrested phase where metallic phase reappears below $T_{MI}$. The kinetically arrested phase is detectable at all the THz frequencies shown by the shaded region in (c) and (d).

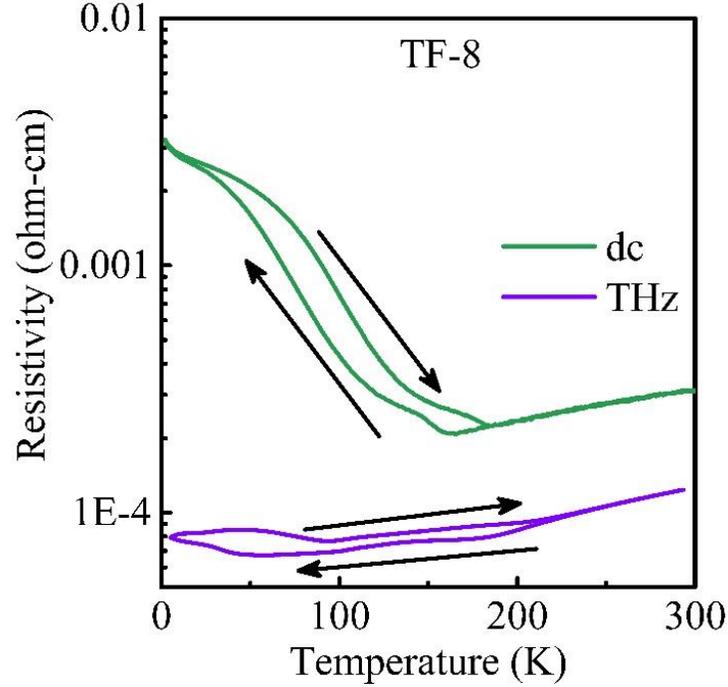

**Figure 4.** Comparative representation of resistivities for TF-8 sample, obtained through DC transport and THz-TDS at 1 THz. Arrows show the path followed by data-points in heating and cooling sequences.

**Table 1:** Description of $RNiO_3$ films used in DC transport and THz-TDS measurement. $T_{MI}$ were determined via DC transport.

|  | **Thin Films** | **Types of Disorder** | **$T_{MI}$** |
|---|---|---|---|
| **Set 1** | TF-1 | Oxygen vacancies | 130 K |
|  | TF-2 | Oxygen vacancies + Cation disorder | 180 K |
|  | TF-3 | Oxygen vacancies + Cation disorder | 150 K |
| **Set 2** | TF-4 | Oxygen vacancies | No MIT |
|  | TF-5 | Oxygen vacancies + Cation disorder | No MIT |
| **Set 3** | TF-6 | Oxygen vacancies + Cation disorder | 209 K |
|  | TF-7 | Oxygen vacancies + Cation disorder | 190 K |
|  | TF-8 | Oxygen vacancies + Cation disorder + Mosaic disorder | 150 K |

AUTHOR INFORMATION

**Corresponding Author**

*E-mail: dsrana@iiserb.ac.in

**Author Contributions**

The manuscript was written through contributions of all authors. All authors have given approval to the final version of the manuscript.

‡These authors contributed equally.

ACKNOWLEDGMENT

Authors acknowledge Prof. Richard Averitt for fruitful discussions. D.S.R. thanks the Science and Engineering Research Board (SERB), Department of Science and Technology, New Delhi, for financial support under research Project No. EMR/2016/003598. Financial support from DST-FIST and CSIR (File No. 09/1020 (0090)/2016-EMR-I) are also thankfully acknowledged.